\documentstyle[12pt]{article}

\catcode`\@=11
\long\def\@makefntext#1{
\protect\noindent \hbox to 3.2pt {\hskip-.9pt  
$^{{\ninerm\@thefnmark}}$\hfil}#1\hfill}                

\def\@makefnmark{\hbox to 0pt{$^{\@thefnmark}$\hss}}  
        
\def\ps@myheadings{\let\@mkboth\@gobbletwo
\def\@oddhead{\hbox{}
\rightmark\hfil\ninerm\thepage}   
\def\@oddfoot{}\def\@evenhead{\ninerm\thepage\hfil
\leftmark\hbox{}}\def\@evenfoot{}
\def\sectionmark##1{}\def\subsectionmark##1{}}

\renewcommand{\thefootnote}{\fnsymbol{footnote}}

\newcounter{sectionc}\newcounter{subsectionc}\newcounter{subsubsectionc}
\renewcommand{\section}[1] {\vspace*{0.6cm}\addtocounter{sectionc}{1} 
\setcounter{subsectionc}{0}\setcounter{subsubsectionc}{0}\noindent 
        {\normalsize\bf\thesectionc. #1}\par\vspace*{0.4cm}}
\renewcommand{\subsection}[1] {\vspace*{0.6cm}\addtocounter{subsectionc}{1} 
        \setcounter{subsubsectionc}{0}\noindent 
        {\normalsize\it\thesectionc.\thesubsectionc. #1}\par\vspace*{0.4cm}}
\renewcommand{\subsubsection}[1]
{\vspace*{0.6cm}\addtocounter{subsubsectionc}{1}
        \noindent {\normalsize\rm\thesectionc.\thesubsectionc.\thesubsubsectionc. 
        #1}\par\vspace*{0.4cm}}

\newcounter{appendixc}
\newcounter{subappendixc}[appendixc]
\newcounter{subsubappendixc}[subappendixc]

\renewcommand{\appendix}[1] {\vspace*{0.6cm}
        \refstepcounter{appendixc}
        \setcounter{figure}{0}
        \setcounter{table}{0}
        \setcounter{equation}{0}
        \renewcommand{\thefigure}{\Alph{appendixc}.\arabic{figure}}
        \renewcommand{\thetable}{\Alph{appendixc}.\arabic{table}}
        \renewcommand{\theappendixc}{\Alph{appendixc}}
        \renewcommand{\theequation}{\Alph{appendixc}.\arabic{equation}}
        \noindent{\bf Appendix \theappendixc #1}\par\vspace*{0.4cm}}

\def\abstracts#1{{
        \centering{\begin{minipage}{12.2truecm}\footnotesize\baselineskip=12pt\noindent
        \centerline{\footnotesize ABSTRACT}\vspace*{0.3cm}
        \parindent=0pt #1
        \end{minipage}}\par}} 

\renewenvironment{thebibliography}[1]
        {\begin{list}{\arabic{enumi}.}
        {\usecounter{enumi}\setlength{\parsep}{0pt}
\setlength{\leftmargin 1.25cm}{\rightmargin 0pt}
         \setlength{\itemsep}{0pt} \settowidth
        {\labelwidth}{#1.}\sloppy}}{\end{list}}

\newcounter{itemlistc}
\newcounter{romanlistc}
\newcounter{alphlistc}
\newcounter{arabiclistc}

\newcommand{\fcaption}[1]{
        \refstepcounter{figure}
        \setbox\@tempboxa = \hbox{\footnotesize Fig.~\thefigure. #1}
        \ifdim \wd\@tempboxa > 6in
           {\begin{center}
        \parbox{6in}{\footnotesize\baselineskip=12pt Fig.~\thefigure. #1}
            \end{center}}
        \else
             {\begin{center}
             {\footnotesize Fig.~\thefigure. #1}
              \end{center}}
        \fi}

\newcommand{\tcaption}[1]{
        \refstepcounter{table}
        \setbox\@tempboxa = \hbox{\footnotesize Table~\thetable. #1}
        \ifdim \wd\@tempboxa > 6in
           {\begin{center}
        \parbox{6in}{\footnotesize\baselineskip=12pt Table~\thetable. #1}
            \end{center}}
        \else
             {\begin{center}
             {\footnotesize Table~\thetable. #1}
              \end{center}}
        \fi}

\def\@citex[#1]#2{\if@filesw\immediate\write\@auxout
        {\string\citation{#2}}\fi
\def\@citea{}\@cite{\@for\@citeb:=#2\do
        {\@citea\def\@citea{,}\@ifundefined
        {b@\@citeb}{{\bf ?}\@warning
        {Citation `\@citeb' on page \thepage \space undefined}}
        {\csname b@\@citeb\endcsname}}}{#1}}

\newif\if@cghi
\def\cite{\@cghitrue\@ifnextchar [{\@tempswatrue
        \@citex}{\@tempswafalse\@citex[]}}
\def\citelow{\@cghifalse\@ifnextchar [{\@tempswatrue
        \@citex}{\@tempswafalse\@citex[]}}
\def\@cite#1#2{{$\null^{#1}$\if@tempswa\typeout
        {IJCGA warning: optional citation argument 
        ignored: `#2'} \fi}}

 1
 1
 1

\font\ninerm=cmr9

\newcommand{\smallz}{{\scriptscriptstyle Z}} 
\newcommand{\smallw}{{\scriptscriptstyle W}} %
\newcommand{\smallh}{{\scriptscriptstyle H}} %

\newcommand{\mz}{M_\smallz}
\newcommand{\mw}{M_\smallw}
\newcommand{\mh}{M_\smallh}
\newcommand{\mt}{M_t}
\newcommand{\mut}{\mu_t}
\newcommand{\rhoc}{\mbox{$ \hat{\rho}$}}
\newcommand{\drcar}{\mbox{$\Delta \hat{r}$}}
\newcommand{\ft}{\footnotesize}
\newcommand{\sineff}{\mbox{$\sin^2 \theta^{{ lept}}_{{ eff}}$} }
\newcommand{\scur}{\mbox{$\hat{s}^2$}}
\newcommand{\sincur}{\mbox{$\sin^{2}\!\hat{\theta}_{\scriptscriptstyle W} 
                           (\mz)$}}

\newcommand{\dr}{\mbox{$ \Delta r$}}

\newcommand{\drhoc}{\mbox{$ \Delta \hat{\rho}$}}

\newcommand{\amtd}{\mbox{$ O(g^4 M_t^2 /\mw^2) $}}
\newcommand{\amtq}{\mbox{$ O(g^4 M_t^4 /\mw^4) $}}
\newcommand{\gmtq}{\mbox{$ O(G_\mu^2 M_t^4) $}}
\newcommand{\smallr}{{\scriptscriptstyle R}} %
\newcommand{\ew}{electroweak}
\newcommand{\msbar}{\overline{\rm MS}}
\def\lequiv{\raise 0.4ex \hbox{$<$} \kern -0.8 em 
\lower 0.62 ex \hbox{$\sim$}}
\def\gequiv{\raise 0.4ex \hbox{$>$} \kern -0.7 em
 \lower 0.62 ex \hbox{$\sim$}}
\newcommand{\equ}[1]{Eq.~(\ref{#1})}
\newcommand{\eqs}[1]{Eqs.~(\ref{#1})}
\newcommand{\be}{\begin{equation}}
\newcommand{\ee}{\end{equation}}
\newcommand{\bea}{\begin{eqnarray}}
\newcommand{\eea}{\end{eqnarray}}
\renewcommand{\thefootnote}{\fnsymbol{footnote} }

\begin{document}
\begin{flushright}
        \small
        MPI-PhT-97-016\\
        February 1997
\end{flushright}
\vspace{8mm}
\centerline{\normalsize\bf TWO-LOOP HEAVY TOP EFFECTS ON PRECISION}
\baselineskip=16pt
\centerline{\normalsize\bf  OBSERVABLES AND THE HIGGS MASS\footnote{Talk 
given at the Workshop on
"The Higgs puzzle - What can we learn from LEP II, LHC, NLC, and FMC?",
Ringberg Castle, Germany, December 8-13, 1996.
To appear in the Proceedings, ed. B. Kniehl.}}
\centerline{\sc Paolo Gambino}
\baselineskip=13pt
\centerline{\footnotesize\it Max Planck Institut f\"ur Physik,
W. Heisenberg Institut,}
\baselineskip=12pt
\centerline{\footnotesize\it F\"ohringer Ring 6, M\"unchen, D-80805, Germany}
\centerline{\footnotesize E-mail: gambino@mppmu.mpg.de}
\vspace*{0.3cm}
\vspace*{0.9cm}
\abstracts{
The \ew\ corrections induced by a heavy top on the main precision observables 
are now available up to \amtd.
The new results    significantly  reduce  the 
theoretical uncertainty and  have a sizable
impact on the determination of \sineff. We give 
precise predictions for  $\mw$
and \sineff in different renormalization schemes,  
estimate their 
accuracy, and discuss the implications
for the indirect determination of $\mh$. 
From the present 
data for \sineff we obtain  $\mh=127^{+143}_{-71}$ GeV (or $\mh \lequiv
 430$ GeV at 95\% C.L.). 
} 
\normalsize\baselineskip=15pt
\setcounter{footnote}{0}
\renewcommand{\thefootnote}{\alph{footnote}}
\vspace{.6cm}

The latest preliminary data from LEP and SLC presented at Warsaw last 
summer\cite{war}  contain no convincing hint of new physics,
in contrast with the situation just a year ago.  
Since the data are in substantial agreement with the Standard Model (SM), 
we can  try to constrain  the 
mass of the SM Higgs boson as much  as possible. 
This is even more  interesting in view of the  Higgs discovery 
potential of the LEP2 program currently under way. 
Indeed,  it has been recently observed \cite{war,hollik}
that at the present level of experimental accuracy the theoretical error 
arising from unknown higher order effects, estimated from  scheme dependence,
has a relevant impact on the indirect bounds on the Higgs 
mass. In the past year  substantial progress has been made concerning the 
higher order corrections of electroweak origin, which is also of relevance 
for the indirect determination of $\mh$. 

In this talk I will briefly illustrate the calculation
of all two-loop electroweak contributions to the main precision observables
which are enhanced by powers  of the top mass; then I will 
discuss the residual scheme and scale dependence of the predictions
when the new results are implemented in different frameworks. 
 After a short  summary of  the 
various sources of theoretical error involved in the calculation of $\mw$ and 
the effective sine, 
I will conclude with a discussion of how the new results
affect the indirect determination of $\mh$.

\section{Calculation of $O(g^4 \mt^2/\mw^2)$ effects}

The very precise measurements carried out at LEP, SLC, and the Tevatron
 in the recent past
have made the study of higher order radiative corrections 
necessary in order to test the Standard Model, 
and possibly to  uncover hints of new physics.
 The one-loop corrections to all the relevant
\ew\ observables are by now 
very well established \cite{YB},
and  two and three-loop effects have been investigated in several  cases. 
The dominance of a heavy top quark in the one-loop \ew\ corrections, 
which depend quadratically on the top mass, has allowed to predict
with good approximation
the  mass of the heaviest quark before its actual discovery. 

Among the 
higher order effects connected with these large non-decoupling
contributions, the QCD corrections are now known 
through $O(\alpha_s^2)$ \cite{QCD}.
As for the  purely \ew\
effects originated at higher orders by the large Yukawa coupling 
of the top,
reducible contributions  have been first studied by Consoli {\it et al.} 
\cite{CHJ}, while a  thorough investigation of leading irreducible 
two-loop contributions has been performed by Barbieri {\em et al.} and others
for arbitrary $\mh$ \cite{barb}.

The result of the calculation of the leading \gmtq\ effects 
on the $\rho$
parameter \cite{barb}  is shown in Fig.1 (upper curve). The 
correction  is relatively sizable
and in the  heavy Higgs case 
 reaches the permille level in the prediction of $\sineff$, 
comparable to the present experimental accuracy \cite{war}.
 \begin{figure}
\input{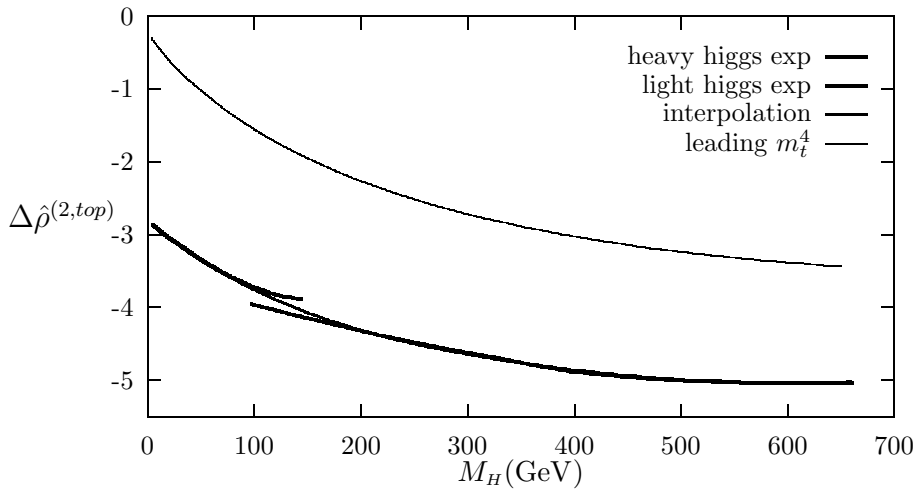}
\vspace{-.6cm}
\caption{\small\sf Two-loop heavy top corrections to $\drhoc$ in 
units $10^{-4}$.  The upper  line 
represents the leading $O(G_\mu^2 m_t^4)$ correction$^6$.}
\vspace{-4mm}
\end{figure}
We observe that the correction is extremely small for a small Higgs mass, 
due to large cancellations. One  can naively expect that setting 
the masses of the vector boson different from zero (and so going beyond 
the pure Yukawa theory considered in \cite{barb}) would spoil the 
cancellations and lead to relevant deviations from the upper curve of 
Fig.1 in the light Higgs region. 

In addition,   the theoretical  uncertainty coming from unknown
higher order effects is dominated by terms \amtd\ \cite{YB}. 
  Indeed, the  renormalization 
scheme ambiguities and the different resummation options examined in
\cite{YB} led to an estimate of the uncertainty of the theoretical 
predictions which was in a few cases  disturbingly sizable, i.e.
comparable to the present experimental error.
In particular, for $\sineff$, the estimated uncertainty was 
$\delta\,\sineff (th)\lequiv\, 1.4\times 10^{-4}$, 
comparable to the present experimental error, $\approx 3\times 10^{-4}$.
For $\mw$ the uncertainty, $\delta{\mw} (th)\lequiv\,
16$ MeV,  was much  smaller than the  error 
on the present world average, 125 MeV.
 A different analysis based on the explicit two-loop calculation 
of the $\rho$ parameter in low-energy processes has also reached 
very similar conclusions \cite{us}.

Motivated  by the previous observations, the complete 
analytic calculation of the 
two-loop quadratic top effects has been performed  
for the relation between the 
vector boson masses and the muon decay constant $G_\mu$ \cite{physlett}
and for the effective leptonic mixing angle, $\sineff$ \cite{zako,dgs}.
 Corrections to the 
 $Z^0$ decay width are also under study.
Using $\alpha$, $G_\mu$, and $\mz$ as inputs, it is then possible to predict 
at this level of accuracy the mass of the $W$ boson and the effective
sine, \sineff. These are the two most accurately known observables at the 
moment, and the ones which allow for the cleanest theoretical interpretation.
The present world averages \cite{war}
 for the $W$ mass and for the effective sine
are $\mw= 80.356\pm0.125$GeV and $\sineff=0.23165\pm0.00024$, respectively.
 The accuracy of their measurement is also going to
 improve  in the near future \cite{baur},
especially in the case of the $W$ mass, 
thanks to LEP2 and the upgrade of the Tevatron.
The precise determination of \sincur,
a by-product of the $\msbar$ calculation, is also important for GUTs studies.

The gauge sector of the SM is described in terms of just three input
quantities, which are routinely chosen to be $\alpha$, the Fermi constant 
$G_\mu$ and $\mz$.
 At any order in perturbation theory we can relate the parameters of the 
Lagrangian to the measured inputs, calculate them, and then make predictions
for any observable. Using 
$s\equiv \sin\theta_\smallw$ and $c\equiv \cos\theta_\smallw$,
we can relate the weak couplings to the inputs through the following two 
relations:
\be
\ \ s^2= 
\frac{\pi\alpha}{\sqrt{2}G_\mu \mw^2} \hspace{1.5cm}\rightarrow
 \hspace{1.3cm}
\ s^2_\smallr= 
\frac{\pi\alpha}{\sqrt{2}G_\mu \mw^2} \frac1{\left(1-
\dr_\smallw^\smallr\right)}\label{drho1}
\ee
\be
\mw^2= c^2 \ \mz^2
\hspace{1.3cm} \rightarrow  \hspace{1.3cm}
\mw^2= c^2_\smallr \ \mz^2 \ \frac{1}{1-\Delta\rho_\smallr}.
\label{drho2}
\ee
Here the l.h.s. corresponds to a tree level description, 
while on the r.h.s. I have shown 
that, in a given renormalization scheme "R", the quantum 
effects can be incorporated through the radiative corrections
$\dr_\smallw^\smallr$ and $\Delta\rho_\smallr$. 
They are functions of $s_\smallr$, the renormalized mixing angle,
and of the physical masses $\mw$, $\mt,\ \mh$, etc.,
and clearly depend on the renormalization scheme.

All the calculations of the \amtd\ effects have been performed in 
the $\msbar$ scheme introduced in \cite{msbar}, i.e. using $\msbar$ 
couplings and on-shell masses, a  particularly convenient framework. 
The mixing angle is then defined by 
\be
\scur\equiv
\sin^2\hat{\theta}_\smallw (\mz)_{\msbar}\equiv
\frac{\hat{\alpha}(\mz)_{\msbar}}{\hat{\alpha}_2(\mz)_{\msbar}}
\label{msbarsine}
\ee
($\hat{\alpha}$ and $\hat{\alpha}_2$ are the $\msbar$ U(1)$_{e.m.}$ 
and SU(2) running couplings),
while the vector boson masses are defined as the physical masses.
In this framework $\Delta\rhoc$ and $\drcar_\smallw$
are the radiative corrections entering the r.h.s. of \eqs{drho1} and 
(\ref{drho2}).
By solving these two equations simultaneously, 
the mass of the $W$ boson and the $\msbar$ mixing angle can be determined.

For what concerns the relation between the effective leptonic sine measured
at LEP and SLC  from  on-resonance asymmetries and the renormalized mixing 
angle, in an arbitrary  
renormalization scheme we find 
\be
\sineff =  \ \left[1+\Delta\, k_{\smallr} (\mz^2)\right] \ s^2_\smallr.
\ee
where $\Delta k_{\smallr} (\mz^2)$ is the real part of an \ew\ form factor 
evaluated at $q^2=\mz^2$. In the $\msbar$ framework  the relevant 
quantity is  $\Delta\hat{k}(\mz^2)$ \cite{rel}.

In this $\msbar$ framework, only $\drhoc$ contains the leading heavy top 
dependence of the \ew\ amplitudes at one and two-loop level,
 i.e. $O(g^2\mt^2/\mw^2)$
and \amtq\ respectively.
 The calculation of the next-to-leading heavy top 
corrections relevant for $\mw$ and \sineff\ therefore consists of computing 
(a) the first corrections to the result of \cite{barb} for $\drhoc$, 
which is displayed in Fig.1; this 
involves only the calculation of on-shell mass counterterms for the 
vector bosons,  and leads to  a striking deviation from the leading result.
(b) the leading \amtd\  contributions to $\drcar_\smallw$ and $\Delta\hat{k}$,
which implies the two-loop 
calculation of the $Z^0$ and the muon decays at \amtd. 

The main tools needed to perform such calculations are:
\begin{itemize}
\item A consistent heavy mass expansion procedure \cite{asym,prep},
whereby the coefficients of the heavy top expansion of any two-loop diagram
are expressed in terms of vacuum two-loop integrals and products of one-loop
 integrals in $n$ dimensions.
\item Vacuum two-loop integrals in $n$ dimensions with arbitrary masses. 
\cite{DT}
\item An efficient algebraic computer package. \cite{PD} 
\item A renormalization procedure kept to  maximal simplicity. The $\msbar$
choice has the advantage that the one-loop coupling conterterms
are not enhanced by powers of $\mt$; only mass renormalization introduces
$\mt^2$ terms. We used bare gauge fixing, which avoids a vector boson--scalar 
mixing counterterm and  has been proved  equivalent to the standard 
procedure\cite{fix}. The Ward identities fix the mass renormalization of the 
unphysical scalars and allow an important check of the vector boson 
self-energies at $q^2=0$.
\end{itemize}
\renewcommand{\arraystretch}{1.1}
\begin{table}[t]
\tcaption{Predicted values of $\mw$ and \sineff in different frameworks, 
including only the leading  $O(g^4 M_t^4)$ irreducible contribution.
QCD corrections based on pole top-mass parametrization. $M_t= 175$GeV.
 }\label{tab1}
\[
\begin{array}{|c|  c c c c|cccc|}\hline
 & \multicolumn{4}{|c|}{\sin^2 \theta_{eff}^{lept}} 
& \multicolumn{4}{|c|}{\mw ({\rm GeV})}\\\hline 
M_H  &  {\rm OSI} & {\rm OSII} 
& \rm{\overline{MS}} & ZFitter
&  {\rm OSI} & {\rm OSII} & \rm{\overline{MS}} & ZFitter
  \\  \hline\hline
65  &  .23131 & .23111 & .23122 &.23116 &80.411 & 80.422 & 80.420 & 80.420 \\ 
100  & .23153 & .23135 & .23144 & .23138& 80.388 & 80.397 & 80.396 & 80.396\\ 
300  & .23212 & .23203 & .23203 & .23197& 80.312 & 80.316 & 80.319 & 80.320\\ 
600  & .23251 & .23249 & .23243 & .23236& 80.256 & 80.257 & 80.263 & 80.265\\
1000 & .23280 & .23282 & .23272 & .23264& 80.215 & 80.213 & 80.221 & 80.224 \\
\hline
\end{array}            
\]
\end{table}

Analytic formulae for the \amtd\ contributions to
$\drhoc$ and $\drcar_\smallw$ can be found in \cite{physlett},
 and in \cite{dgs}
for the corresponding 
$\Delta\hat{k}(\mz^2)$.
The details of the calculation are presented elsewhere \cite{zako,prep}.
An interesting comparison \cite{weiglein}
between numerical evaluation and heavy top expansion
of  two-loop diagrams has been made  for a few cases 
of relevance in this calculation. It shows a very good convergence of 
the expansion, already acceptable for $\mt=175$GeV.
The numerical calculation\cite{weiglein}
of the two-loop Higgs mass dependent top contributions  
to $\dr$ is also in very good agreement with our results,
within the error range to be discussed in the following.

The QCD corrections play an important role in \ew\ calculations \cite{QCD}
 and are intertwined with the pure \ew\ contributions in a subtle way.
The most important effect is a substantial (about 12\%)
screening of the leading top contribution 
to $\drhoc$ when this is expressed in terms of the top pole mass $M_t$. 
The coefficients of the $a\equiv\alpha_s/\pi$ series are then 
large and increasing with the power of $a$,
 a feature that is absent  when the amplitudes
are expressed in terms of the $\msbar$  mass $\mut= \hat{m}_t
(\mut)$. The leading $m_t^2$ term is rescaled by a factor $(1+\delta_{QCD})$
which in the pole mass and $\mut$ parametrizations is given by, respectively:
\be
\delta_{QCD}^{\mt}= -2.8599 \,a(\mt) - 14.594\, a^2(\mt); \ \ \ 
\delta_{QCD}^{\mut}= -0.19325\, a(\mut) - 3.970\, a^2(\mut).
\label{venti}
\ee
Physically, this can be attributed to the fact that the use of the quark pole 
mass  introduces a spurious 
sensitivity to long distance dynamics, witnessed by the appearance of 
the leading IR renormalon in the QCD perturbative expansion\cite{sirqcd}.
The natural expansion parameter for \ew\ physics is therefore  a high-scale
mass, such as $\mut$. This can be  derived from the measured $\mt$ 
by following the strategy outlined 
in the first paper of \cite{sirqcd}  and optimizing the perturbative series.
 For our numerical study we implement 
QCD corrections up to  $O(\alpha\alpha_s^2)$ adopting 
consistently either the $M_t$ or the $\mut$ parametrizations and  using 
 $\mz=91.1863$ GeV, 
$\alpha_s(\mz)=0.118$, $\Delta\alpha_{had}= 0.0280$,
$\mt=175$ GeV as inputs. For each of the 
top mass parametrizations, 
the predictions for $\mw$ and $\sineff$ 
 in the $\msbar$ scheme are shown  in the corresponding  columns of
Table 2 and 3 for different $\mh$ values.
 Table 1 shows the pole mass result in the case the new \amtd\
corrections to $\drhoc$ and $\drcar_\smallw$ are neglected. Results
for $\sincur$ can be found in \cite{dgs}.
\begin{table}[t]
\tcaption{As in Table 1, but including the new $O(g^4 M_t^2)$ 
contributions. }\label{tab2}
\[
\begin{array}{|c|  c c c| c c c|}\hline
 & \multicolumn{3}{|c|}{\sin^2 \theta_{eff}^{lept}} 
& \multicolumn{3}{|c|}{\mw ({\rm GeV})}\\\hline 
M_H  &  {\rm OSI} & {\rm OSII} & \rm{\overline{MS}}&  {\rm OSI} & {\rm OSII} 
& \rm{\overline{MS}} 
  \\  \hline\hline
65  &  .23132 & .23134 & .23130  &  80.405 & 80.404 & 80.406     \\ 
100  & .23153 & .23155 & .23152   & 80.382 & 80.381 & 80.383  \\ 
300  & .23210 & .23214 & .23210 & 80.308 & 80.306 & 80.308\\ 
600  & .23249 & .23252 & .23249  & 80.254 & 80.252 & 80.254\\ 
1000 & .23277 & .23279 & .23277 & 80.214 & 80.213 & 80.214 \\ \hline 
\end{array}            
\]
\end{table}

\section{Scale and scheme dependence}
\begin{figure}[t]
\input{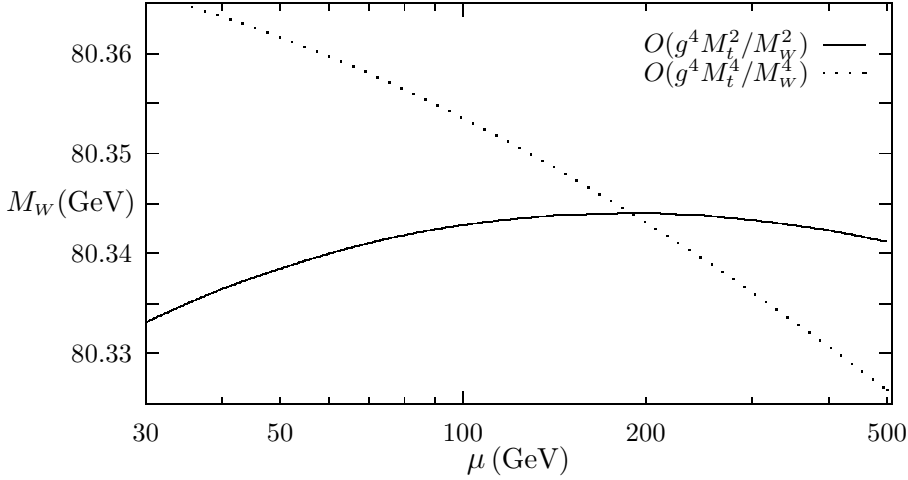}
\vspace{-6mm}
\caption{\small\sf Scale dependence of the prediction for $\mw$ 
in the $\msbar$
scheme for $\mt=180$GeV, $\mh=300$GeV, including only the leading
\gmtq\ correction (dotted curve)
or all the available two-loop contributions, through
\amtd\ (solid curve).}
\label{mudep}
\vspace{-3mm}
\end{figure}

In the cases of $Z^0$ and muon decays,
 the natural scale for  the $\msbar$ couplings is 
of the order of the vector boson masses.
 Conventionally \cite{msbar}  one   sets
$\mu=\mz$; however, one may also consider the case of a general $\mu$.
In that case, although  physical observables are  $\mu$-independent,
a residual scale dependence is left in the $\msbar$ calculation of 
$\mw$ and \sineff\ in $O(g^4)$. Of course, the $\mu$-independence of
the \amtd\ terms can be explicitly verified.
The scale dependence of our predictions can therefore be used to gauge the
importance of  uncalculated higher order effects.
The situation is 
exemplified in Fig.\ref{mudep}. 
In the range 50GeV$<\mu<$500GeV, $\mw$ varies by less than 5MeV. 
Over a wide range of $\mu$ values, the
scale dependence of  $\mw$ 
is  significantly reduced by the inclusion
of the \amtd\ contribution. 
The case of \sineff\ is very similar.

In order to discuss the scheme dependence of our predictions, 
it is convenient  to combine \eqs{drho1} and (\ref{drho2})
and  calculate
directly the $W$ boson mass from $\alpha$, $\mz$, and $G_\mu$:
\be
\frac{\mw^2}{\mz^2} \left( 1-\frac{\mw^2}{\mz^2}\right)=
\frac{\pi \alpha}{\sqrt{2} G_\mu\mz^2} \frac1{1-\dr}
\label{deltar}
\ee
where, in the $\msbar$ scheme, using $s^2=1-c^2$ and $c= \mw/\mz$,
\be
1-\dr= \left[1-\drcar_\smallw\right] \left(1+ \frac{c^2}{s^2} \drhoc\right).
\label{drms}
\ee
Unlike $\drhoc$ and $\drcar_\smallw$, $\dr$  
is a physical observable.
In the on-shell (OS) renormalization scheme of \cite{si80} the couplings
are defined directly by the mass relation \equ{drho2}, where $\Delta\rho_{\sf os}\equiv0$, $c^2_{\sf os}\equiv c^2=\mw^2/\mz^2$,
  and by the traditional QED charge renormalization.
Starting from \equ{drms} it is easy  to translate our results into the OS
framework. We notice that $\drhoc$ and $\drcar_\smallw$ are functions,
respectively, 
of the  couplings $\scur$, $\hat{e}^2$, and $\scur$, $e^2=4\pi \alpha$;
$\scur$ is  related to the 
OS counterparts precisely by the $\msbar$ versions of 
Eqs.(\ref{drho1},\ref{drho2}). 
Replacing $\msbar$ couplings with OS ones in a systematic way
\cite{dgs}, one obtains
\bea
\Delta\bar{r}_\smallw(s) &=& \drcar_\smallw^{(1)}(s^2) + 
\drcar_\smallw^{(2)}(s^2) +\Delta\bar{r}_\smallw^{(2,add)}(s^2)\nonumber\\
\Delta\bar{\rho}(s)&=& \drhoc^{(1)}(s^2) + \drhoc^{(2)}(s^2)+
\Delta\bar{\rho}^{(2,add)}(s^2)
\eea
where the indices (1,2) indicate the one and two-loop part of the corresponding
functions. The one loop contributions can be gleaned from \cite{msbar}, while
the additional 
two-loop contributions are explicitly given in \cite{physlett,dgs}.
It is  important to notice that $\Delta\bar{r}_\smallw$ must be 
expressed in terms of $\alpha$, and $\Delta\bar{\rho}$ in terms of $G_\mu$.
\equ{drms} can now be rewritten  in the OS scheme as
\be
1-\dr= \left[1-\Delta\bar{r}_\smallw\right] \left(1+ \frac{c^2}{s^2} 
\Delta\bar{\rho}\right).
\label{dros}
\ee
This equation is similar to the one obtained in \cite{CHJ}, but
fixes the correct resummation of subleading effects at \amtd.
We refer to the framework of \equ{dros} as OSI. After deriving in a similar
 way the corrections relevant for \sineff\ (the one-loop part is given in
\cite{rel}, the additional two-loop are given again in \cite{dgs}),
 we obtain the predictions shown in
Tables 1,2,3 in the corresponding column.  

OSI is a hybrid scheme,
which presupposes a $\msbar$ subtraction
and retains a  residual $\mu$-dependence in  $\dr$.
An alternative OS resummation can be obtained expanding \equ{dros} in powers of
$\mt$, keeping terms only up to \amtd. With the understanding that
everything is expressed in terms of $\alpha$, the result is
\be
\Delta r = \Delta r^{(1)}
 + \Delta r^{(2)}  
  + \left( \frac{c^2}{s^2}\right)^2 
N_c\, x_t \left( 2 \Delta\bar{\rho}^{(1)}(s^2) -
N_c\frac{\alpha}{16\pi \,s^2} \frac{\mt^2}{\mw^2}\right)
\label{quindici}.
\ee
where $\dr^{(1)}$ is the original one-loop OS result of \cite{si80}
expressed in terms of $\alpha$ and $s^2$, $\dr^{(2)}=
\Delta\bar{r}_\smallw^{(2)} - (c^2/s^2)\, \Delta\bar{\rho}^{(2)}$,
$N_c x_t$ is the leading $\mt^2$ part of $\drhoc^{(1)}$, 
and $\Delta\alpha$ is 
the renormalized photon
vacuum polarization function at $q^2=\mz^2$. We refer to this OS framework 
as OSII. It retains no record of the $\msbar$ derivation.
The results are  shown again in Tables 1,2,3.
For comparison purposes I also show in Table 1 the predictions computed by 
{\it ZFitter} \cite{zfitter}, 
where the new contributions are not yet implemented. 
\begin{table}[t]
\tcaption{As in Table 2, but parametrizing the top mass in terms of $\mut$.
 }\label{tab3}
\[
\begin{array}{|c|  c c c| c c c|}\hline
 & \multicolumn{3}{|c|}{\sin^2 \theta_{eff}^{lept}} 
& \multicolumn{3}{|c|}{\mw ({\rm GeV})}\\\hline 
M_H  &  {\rm OSI} & {\rm OSII} & \rm{\overline{MS}}&  {\rm OSI} & {\rm OSII} 
& \rm{\overline{MS}} 
  \\  \hline\hline
65  &  .23132 & .23132 & .23130  &  80.404 & 80.404 & 80.406     \\ 
100  & .23152 & .23154 & .23151   & 80.381 & 80.381 & 80.383  \\ 
300  & .23209 & .23212 & .23209 & 80.308 & 80.307 & 80.309\\ 
600  & .23248 & .23250 & .23247  & 80.255 & 80.254 & 80.256\\ 
1000 & .23275 & .23277 & .23275 & 80.216 & 80.215 & 80.216 \\ \hline 
\end{array}            
\]
\end{table}

We can now compare the results in the three approaches. They share the same
virtues and are  equivalent up to 
$O(g^4)$ terms not enhanced by power of $\mt$ or
by large logarithms $\ln m_f/\mz$. 
 As expected, the scheme dependence
of the predicted values of $\sineff$ and $\mw$ is drastically reduced
by the inclusion of the \amtd\ contributions, by a factor 
comparable to the expansion parameter $\mw^2/\mt^2\approx 0.2$.
For instance, the maximum differences in Table 1 are 11 MeV in $\mw$ and
$2.1\times 10^{-4} $ in $\sineff$, while in Table 2 they are reduced 
to 2 MeV and $4\times 10^{-5} $, respectively (similar small 
differences can be observed in Table 3). 
 Although the QCD approaches we have considered are quite different,
Tables 2 and 3 show very close results. This is due to a curious 
cancellation of screening and anti-screening effects 
in the difference between the two formulations.
 Because of the sign of the shifts, in general the \amtd\ correction 
further enhances the screening of the top quark
 contribution by higher order effects. 

By comparing the results in Tables 2 and 3 
we  estimate the error attached to 
our predictions. 
As there is   very close and accidental 
agreement between Tables 2 and 3, we can use  scale dependence\cite{bernd} 
or other  methods\cite{sirqcd}
to pin down the irreducible uncertainty coming from QCD corrections.
The values given in Table 4 are somewhat conservative, 
and do {\it not} correspond to half-differences (see Table 2).
In Table 4 I also display
the main parametric uncertainties, i.e. the ones due to the 
inaccuracy of the input parameters. The dominant effect on \sineff
is connected to the 
evolution of the e.m. coupling from $q^2=0$ up to the weak scale, 
which involves long-distance dynamics.
The uncertainty due to variations of $\mh$ between 65GeV and 1TeV
measures  the sensitivity of the observables to
the Higgs boson mass. 
We can conclude that  $\dr$ can now be determined 
with high  theoretical accuracy, at the level     of   $1\times 
10^{-4}$, as the theoretical
error coming from higher order contributions appears to be well under control.
The improvement due to the new calculation of \amtd\ effects is pictorially 
evident in Fig.3, where the predictions for \sineff\ in different schemes
are compared when \amtd\ effects are included or not. 
The dominant parametric uncertainty is also shown.
\begin{table}[t]
\tcaption{Parametric and intrinsic uncertainties in the calculation of $\mw$ 
and $\sineff$.}\label{tab4}
\[
\begin{array}{|c|  c| c |}\hline
{\rm source \ of \ uncertainty}
 & \delta\mw &\delta\sin^2 \theta_{eff}^{lept} \\\hline \hline
\delta\mt=6\ {\rm GeV} & 36\ {\rm MeV} & 1.9 \times 10^{-4}   \\ 
 \delta\,\alpha_s(\mz)=0.005 & 3\ {\rm MeV} & 2 \times 10^{-5}   \\ 
 \delta\mz=2 \ {\rm MeV} & 2\ {\rm MeV} & 2 \times 10^{-5}   \\ 
\delta\alpha(\mz)/\alpha(\mz)=7\times 10^{-4} & 13\ {\rm MeV} & 2.3 \times 10^{-4}   \\ 
\mh=65\,-\, 1000 {\rm GeV} & 190\ {\rm MeV} & 14  \times 10^{-4}   \\
{\rm higher\ order\ contr.\ EW} & 2\ {\rm MeV} & 4 \times 10^{-5}   \\ 
{\rm higher\ order\ contr.\ QCD} & 5\ {\rm MeV} & 3 \times 10^{-5}   \\ \hline
\end{array}            
\]
\end{table}

\section{Indirect Higgs mass determination \cite{dgs}}
Unlike the case of the top quark, the dependence of \ew\ amplitudes on
the Higgs mass is very mild, at most logarithmic at one-loop level.
This explains the very loose bounds that even the present very accurate data
 put on $\mh$. 
With the present experimental errors, 
\sineff\ is the observable most sensitive  to the Higgs boson mass \cite{war}
(see also Table 4).
Most of the resolution on $\mh$ that the global fits show comes from this 
single precise measurement.
However, Fig.3 shows that even if we could reduce the experimental 
error on \sineff\ by a large factor, a strict bound on $\mh$ would be 
hampered by the large hadronic uncertainty. Still, the new \amtd\
results do have 
an impact  in the present situation.
\begin{figure}
\vspace{-2mm}
\input{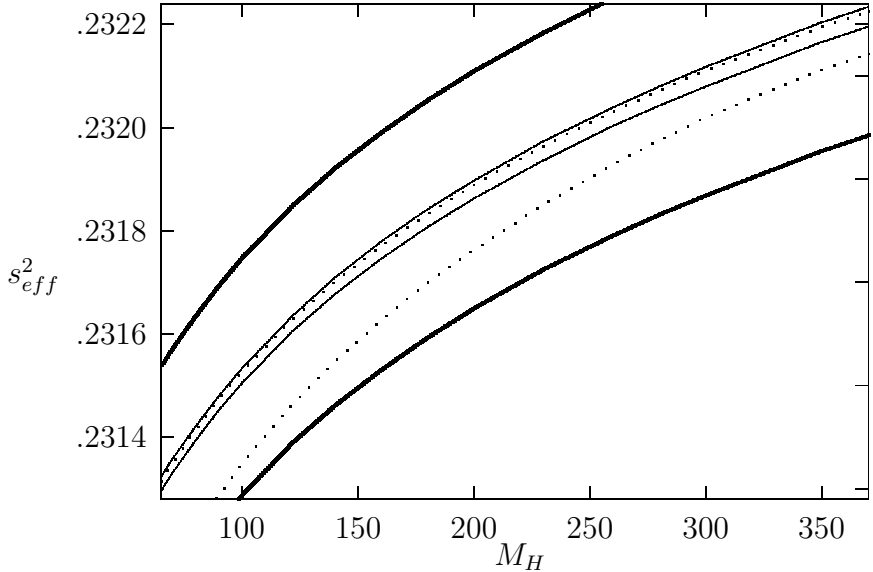}
\vspace{-.6cm}
\caption{\small\sf Predictions for \sineff\ as functions of $\mh$. 
The values between the thin solid (dotted) lines 
identify the predictions in the three schemes described in the text
with (without) the irreducible 
\amtd\ contributions and using either the $\mt$ or $\mut$ 
parametrizations.
The thick solid lines define  the range of values allowed when 
the main parametric 
uncertainty, coming from $\alpha(\mz)$, is included.}
\vspace{-4mm}
\end{figure}

The results of Table 2 and 3 are well described  by 
$\sineff=0.23153 + 5.2\times 10^{-4} \ln(\mh/100{\rm GeV})$.
Comparing the current world average $\sineff=0.23165\pm 0.00024$ with 
the previous expression, and taking into account in quadrature the errors 
 given in 
Table 4, we obtain $ 5.2\times 10^{-4} \ln (\mh/100{\rm GeV})= (1.2\pm 3.9)
\times 10^{-4}$, which implies
 $\mh=127^{+143}_{-71}$GeV, compatible with MSSM expectations. 
This corresponds to an upper bound  $\mh\lequiv 430$GeV at 95\% C.L. (or 
1.64$\sigma$)
and  compares well with the LEP global fit result\cite{war} $\mh=149^{+148}_{
-82}$GeV (or $\mh\lequiv 550$GeV at 95\% C.L. when the theoretical errors
estimated in \cite{YB} are included, shifting the upper bound by $\approx 
+100$GeV).
Despite
the fact that our determination stems from a single piece of data,
it has  reduced errors
and lower central value.
This can be understood by noting that (i) as shown in Fig.3 and Tables 2 and 3,
when the \amtd\ corrections are included, the scheme dependence is 
much smaller than estimated in \cite{YB}.
(ii) the new corrections generally 
enhance  the screening of the one-loop top contribution induced by other 
higher order effects (QCD and leading \amtq\ corrections); as it is clear
from Fig.3,  
the  prediction of  higher \sineff\ values leads to lower central value
 for $\mh$. (iii)   because of the approximate
exponential dependence of $\mh$ on $\sineff$, a small shift in the predicted 
value of the effective sine implies  a significant change in the 
determination of $\mh$.
For instance, a 0.1\% difference in the theoretical calculation of \sineff\
induces a $\approx$55\% shift in the \sineff\ determination of $\mh$ and its
$1\sigma$ bound. In our case higher predicted \sineff\ values imply lower 
$\mh$,  hence smaller $\delta\mh$.
We expect that a global analysis performed including the \amtd\ correction
with the same data
should bring down the upper bound on $\mh$ by a significant amount.

In summary, 
two-loop \ew\ \amtd\ effects are now available
in analytic form
for the main precision observables in {\em both} $\msbar$ and OS 
frameworks. The new contributions consistently reduce the scheme
and scale dependence of the predictions by {\em at least}
a factor $\mt^2/\mw^2\approx 5$, suggesting 
a relevant improvement in the 
theoretical accuracy. The impact on 
the value of \sineff can be sizable, up to 2$\times 10^{-4}$,
depending on   the scheme.
The  enhancement of the screening  of the dominant top contribution
from higher order corrections and the sharp reduction of the  theoretical
uncertainty
lead to  bounds on the Higgs mass from 
present data with reduced errors and lower central value:
  $\mh=127^{+143}_{-71}$GeV or 
$\mh\lequiv 430$GeV at 95\% C.L.

\vspace{.2cm}
 I am grateful to Bernd Kniehl for the excellent organization and
the pleasant atmosphere during the workshop.
I wish to thank my collaborators G. Degrassi, 
A. Sirlin and A. Vicini, with whom the work described here has been done,
and   G. Ganis, W. Hollik, B. Kniehl, 
and G. Weiglein for useful discussions and communications.


\begin{thebibliography}{99}

\bibitem{war} A. Blondel,  talk at ICHEP 96, Warsaw, July 1996,
and the LEP Electroweak Working Group, Internal 
Note LEPEWWG/96-02.

\bibitem{hollik} W. Hollik,  talk at ICHEP 96, Warsaw, July 1996,
hep-ph/9610457.

\bibitem{YB} D. Bardin et al., in {\it 
Reports of the Working Group on Precision 
Calculations for the Z-resonance}, CERN Yellow Report, CERN 95-03, 
D. Bardin, W. Hollik, and G. Passarino eds., p. 7.

\bibitem{QCD} 
A. Djouadi, Nuovo Cimento {\bf 100A} (1988) 357;
B.A. Kniehl, Nucl. Phys. {\bf B347} (1990) 86;
 A. Djouadi, P. Gambino, Phys.\ Rev.\ {\bf D49} (1994) 3499,
 E: {\it ibid.} {\bf D53} (1996) 4111;
K.G. Chetyrkin, J.H. K\"uhn, M. Steinhauser, Phys. Rev. Lett. {\bf 75},
(1995) 3394; L. Avdeev {\em et al.}, Phys. Lett. {\bf B336} (1994) 560, E:
{\it ibid.}  {\bf B349} (1995) 597.

\bibitem{CHJ} M.~Consoli, W.~Hollik, F.~Jegerlehner, Phys.~Lett. {\bf 
    B227} (1989) 167.

\bibitem{barb} R.~Barbieri {\em et al.},
Nucl. Phys. {\bf B409}             (1993) 105;
J.~Fleischer, O.V.~Tarasov, F.~Jegerlehner, Phys.
Lett.     {\bf B319} (1993) 249;
  G.~ Degrassi, S.~Fanchiotti,  P.~Gambino, 
               Int.~J.~Mod.~Phys. {\bf A10} (1995) 1337.

\bibitem{us} G. Degrassi {\em et al.}, 
Phys. Lett. {\bf B350}, (1995) 75.

\bibitem{physlett} G.~Degrassi, P.~Gambino,  A.~Vicini,
Phys. Lett. {\bf B383} (1996) 219.

\bibitem{zako} P. Gambino, 
 hep-ph/9611358, Acta Phys. Pol. 
{\bf B27},  (1996) 3671.

\bibitem{dgs} G.~Degrassi, P.~Gambino, A.~Sirlin, hep-ph/9611363,
Phys. Lett. {\bf B}, to appear.

\bibitem{baur} U. Baur, M. Demarteau, hep-ph/9611334.
\bibitem{msbar} A.~Sirlin, Phys. Lett. {\bf B232}  (1989) 123;
  G.~ Degrassi, S.~Fanchiotti,  A.~Sirlin, Nucl. Phys.
              {\bf B351} (1991) 49.

\bibitem{rel} 
P. Gambino, A. Sirlin, Phys. Rev. {\bf D49} (1994) 1160.

\bibitem{asym} See V.A. Smirnov,  Mod. Phys. Lett.{\bf A10} (1995) 1485,
and refs. therein;
F. Berends, A.T. Davydychev, V.A. Smirnov, Nucl. Phys.{\bf B478} (1996), 59.

\bibitem{prep} G.~Degrassi, P.~Gambino, in preparation.

\bibitem{DT} A.T. Davydychev, B. Tausk, Nucl. Phys. {\bf B397} (1993) 123.
\bibitem{PD} G.~Degrassi, S.~Fanchiotti, and  P.~Gambino, 
{\it ProcessDiagram}.


\bibitem{fix} G. Degrassi, F. Feruglio, private communication.

\bibitem{weiglein} S. Bauberger and G. Weiglein, hep-ph/9611445 and 
private communication.

\bibitem{sirqcd} A. Sirlin, Phys. Lett. {\bf B348}, 201 (1995); P. Gambino 
and A. Sirlin, Phys. Lett. {\bf B355} (1995) 295.
\bibitem{si80} A. Sirlin, Phys. Rev. {\bf D22} (1980) 971.
\bibitem{zfitter} D. Bardin {\it et al.}, {\tt ZFITTER 4.9}, hep-ph/9412201.

\bibitem{bernd} B.A. Kniehl, in Ref.\cite{YB}, p. 299.

\end{thebibliography}
\end{document}